\documentstyle[aps,multicol,epsfig]{revtex}
\begin{document}
\draft
\title{Peak Effect of $J_c$ and Nonlinear $E-J$ Characteristies of High-$T_c$
Superconductors
\thanks{This work is supported by the Ministry of Science
and Technology of China (NKBRSF-G 19990640) and the Chinese
NSF.}}
\author{W.Wang, Z.Qi, H.Y.Xu, F.R.Wang, C.Y.Li and D.L.Yin$^\ast$}
\address{Department of Physics, Peking University, Beijing 100871, China}
\maketitle

\begin{abstract}
The critical current density $J_c$ in high-$T_c$ superconductors
(HTS) often shows a maximum at field far above the self-field. We
study this peak effect (PE) with the nonlinear $I-V$ response of
type-II superconductors and find analytical equation of $J_c$ in
the dependence of field and temperature. This equation is
compared with some experimental data of R-Ba-Cu-O single crystals
with fair agreement.

\vspace{0.5cm}
\end{abstract}


\begin{multicols}{2}
The vortex matter becomes mobile when the driving current density
exceeds some critical value Jc. In high-Tc superconductor RBa$_2$Cu$%
_3O_{7-\delta }$ (R=Y, Gd, Tm and Nd) Jc often shows a maximum at applied
magnetic fields far above the self-field. This peak effect is of great
interest from both fundamental amd technological aspects and studied
extensively \cite{1}-\cite{6}. Scaling behavior in the form $%
J_c(B,T)/J_c(B_p,T)=f(B/B_p)$ with peak field $B_p$ has been observed in
twinned and twin-free YBa$_2$Cu$_3$O$_{7-\delta }$ single crystals \cite{2}.
Prominent PE of $J_c$ with the scaling behavior of the pinning force $F_p$
in the form $F_p/F_{pmax}=f(B/B_{irr})$ has been observed in NdBa$_2$Cu$_3$O$%
_{7-\delta }$ single crystals \cite{3}, where $B_{irr}$ is the
irreversibility field. Gurevich and Vinokur showed that PE could be
interpreted by the nonlinear transport properties of type-II superconductors
with macroscopic inhomogeneities \cite{5}. Otterlo et al. explained PE by
phase transitions between a stiffer and softer vortex phase \cite{6}.
Nevertheless, comparison of models study with pertinent experimental data is
still lacking. In present work we derive the analytical function of $J_c(B)$
from the nonlinear materials equation of HTS and compare it with experiments.

One of the central issues of physics of the mixed state in type-II
superconductors is thermally activated vortex creep characterized by highly
nonlinear electric-current density ($E-J$) characteristics below the
critical current density $J<J_c$. This kind of characteristics is usually
expressed as

\begin{equation}
E(J)=J\rho _fe^{-U(J,B,T)/kT}
\end{equation}

with $\rho_{f}\approx\rho_{n}B/B_{c2}$ the flux-flow resistivity
estimated by Bardeen and Stephen \cite{12}.

Different models used different types of $U(J)$ which are suggested to
approximate the real barrier. For instance, the Anderson-Kim model with $%
U(J)=U_c(1-J/J_{c0})$\cite{7}\cite{8}, the logarithmic barrier $U(J)=U_c\ln
(J_{c0}/J)$\cite{9}, and the inverse power-law $U(J)=U_c\ln [(J_{c0}/J)^\mu
-1]$\cite{11}\cite{12}. It is recently shown \cite{13}\cite{14}, if one
makes a common modification to the different model barriers $U(J)$ as
\begin{equation}
U(J)\longrightarrow U(J_p\equiv J-E(J)/\rho _f)
\end{equation}

The corresponding modified materials equation. (1) leads to a common
nonlinear form as
\begin{equation}
y=x\exp [-\gamma (1+y-x)^p]
\end{equation}

with $x$ and $y$ the normalized current density and electric field
respectively. $\gamma $ is a parameter characterizing the symmetry breaking
of the pinned vortices system and $p$ is an exponent.

In connection with the Anderson-Kim model
\[
E(J)=2\nu _0B\exp [(-U_0-W_V)/kT]\sinh (W_L/kT)
\]

or

\begin{equation}
E(J)=J\rho _f\exp [(-U_0-W_V+W_L)/kT]  \label{EJ}
\end{equation}

We have in Eq. (3) $p=1$, $\gamma =U_0/kT,$ $x=W_V/kT$ and $y=W_L/kT$. $W_V$
is the dissipation energy due to flux moving. From the result of Bardeen and
Stephen\cite{12}:

\begin{equation}
W_V=\eta \cdot v\cdot A=E(J)\cdot B\cdot A/\rho _f  \label{WV}
\end{equation}

with the viscous drag coefficient $\eta =B\cdot B_{c2}/\rho _n=B^2/\rho _f$.
$A$ is the product of the volume of vortex bundles and the range of pinning
force.

$W_L$ is the work done by Lorentz force.

\begin{equation}
W_L=J\cdot B\cdot A  \label{WL}
\end{equation}

The critical current density $J_c$ is defined by a certain criterion $%
E(J_c)\equiv E_c$. From upper equation. $\left( \ref{EJ}\right) $, $\left(
\ref{WV}\right) $ and $\left( \ref{WL}\right) $ we get:

\begin{equation}
J_c=J_{c0}[1-\frac{kT}{U_0}\ln (\frac{\nu _0B}{E_c})+\frac{E_c}{\rho _fJ_{c0}%
}]  \label{jc}
\end{equation}

where$J_{co}$ is the critical current density without the help of thermal
activation.
\begin{equation}
J_{c0}=U_0/BA,  \label{jc0}
\end{equation}
and $\ln (\nu _0B/E_c)$ is a slow varying function of the magnetic field $B$%
; therefore, we set $\ln (\nu _0B/E_c)=\ln (E_0/E_c)$ as a constant. From
Eq. $\left( \ref{jc}\right) $ and according to the criterion about $E_c$,
the irreversible field $B_{irr}$ can be defined as:
\[
U_0(T,B_{irr})=kT\ln (E_0/E_c).
\]
When $B\approx B_{irr}$, $J_c\approx E_c/\rho _f$, the system will turn into
flux flow region. The critical current density can be expressed as
\begin{equation}
J_c=J_{c0}\left[ 1-\frac{U_0(T,B_{irr})}{U_0(T,B)}+\frac{E_cB_{c2}}{\rho
_nBJ_{c0}}\right]  \label{Fpc1}
\end{equation}
We assume that the dependence of the pinning potential $U_0$ on temperature
and field can be separated as
\begin{equation}
U_0(T,B)=\alpha (T)B^l(1-b)^m,\ b\equiv B/B_{c2}.  \label{U0}
\end{equation}
and note that the factor $BA$ in Eq. $\left( \ref{jc0}\right) $ has the form
\cite{15}
\begin{equation}
BA\propto \left[ 1-\frac T{T^{*}(B)}\right] ^{-3\nu }\approx \left[ 1-\zeta
\left( \frac B{B_{irr}}\right) ^{-l}\left( \frac{1-b}{1-b^{*}}\right)
^{-m}\right] ^{-3\nu }  \label{BA}
\end{equation}
where $\zeta \equiv \alpha (T)/\alpha [T^{*}(B)]$ is a slow variable and $%
b^{*}\equiv B_{irr}/B_{c2}$. Substitute Eqs. $\left( \ref{U0},\ref{BA}%
\right) $ into $\left( \ref{Fpc1}\right) $, we have
\linethickness{0.7pt}
\end{multicols}
\begin{picture}(4,5)
\put(-5,4){\line(1,0){90}}
\put(85,3.9){\line(0,1){2}}
\end{picture}

\begin{eqnarray}
J_c &=&\alpha ^{^{\prime }}(T)\left[ 1-\zeta (C-1)^m\left( \frac B{B_{irr}}%
\right) ^{-l}\left( C-\frac B{B_{irr}}\right) ^{-m}\right] ^{-3\nu }\left(
\frac B{B_{irr}}\right) ^l\left( C-\frac B{B_{irr}}\right) ^{-m}  \label{Jc}
\\
&&\times \left[ 1-(C-1)^m\left( \frac B{B_{irr}}\right) ^{-l}\left( C-\frac B%
{B_{irr}}\right) ^{-m}+\frac{E_c}{J_{c0}\rho _n}\frac 1b\right]  \nonumber
\end{eqnarray}

\begin{picture}(4,5)
\put(90,0){\line(1,0){90}}
\put(90,0.2){\line(0,-1){2}}
\end{picture}
\begin{multicols}{2}
where $C\equiv B_{c2}/B_{irr}$ and $\alpha ^{^{\prime }}(T)\equiv
C^{-m}B_{irr}^l\alpha (T)$.
The scaling form of the numerical solutions of Equation $(\ref{Jc})$ are
shown in Fig. 2 with prominent peaks form like that observed in \cite{3}
Fig. 1. The comparison of Eq. (\ref{Jc}) with the experimental data of Ref.
[3] in the scaling form is shown in Fig. 2 where we see a fair agreement.

In summary, we show a common nonlinear electric field-current density ($E-J$%
) characteristics equation. And from this equation we can get a clear
expression of critical current density $J_c$. The widely observed peak
effect can be well understood by this equation. Furthermore, we found that
the peak effect under different temperature can be scaled as well.

This work is supported by the Ministry of Science and Technology of China
(NKBRSF-G19990640) and the Chinese NSF.

\begin{figure}[F1]
\epsfxsize= .83\hsize  \vskip 1.0\baselineskip \centerline{
\epsffile{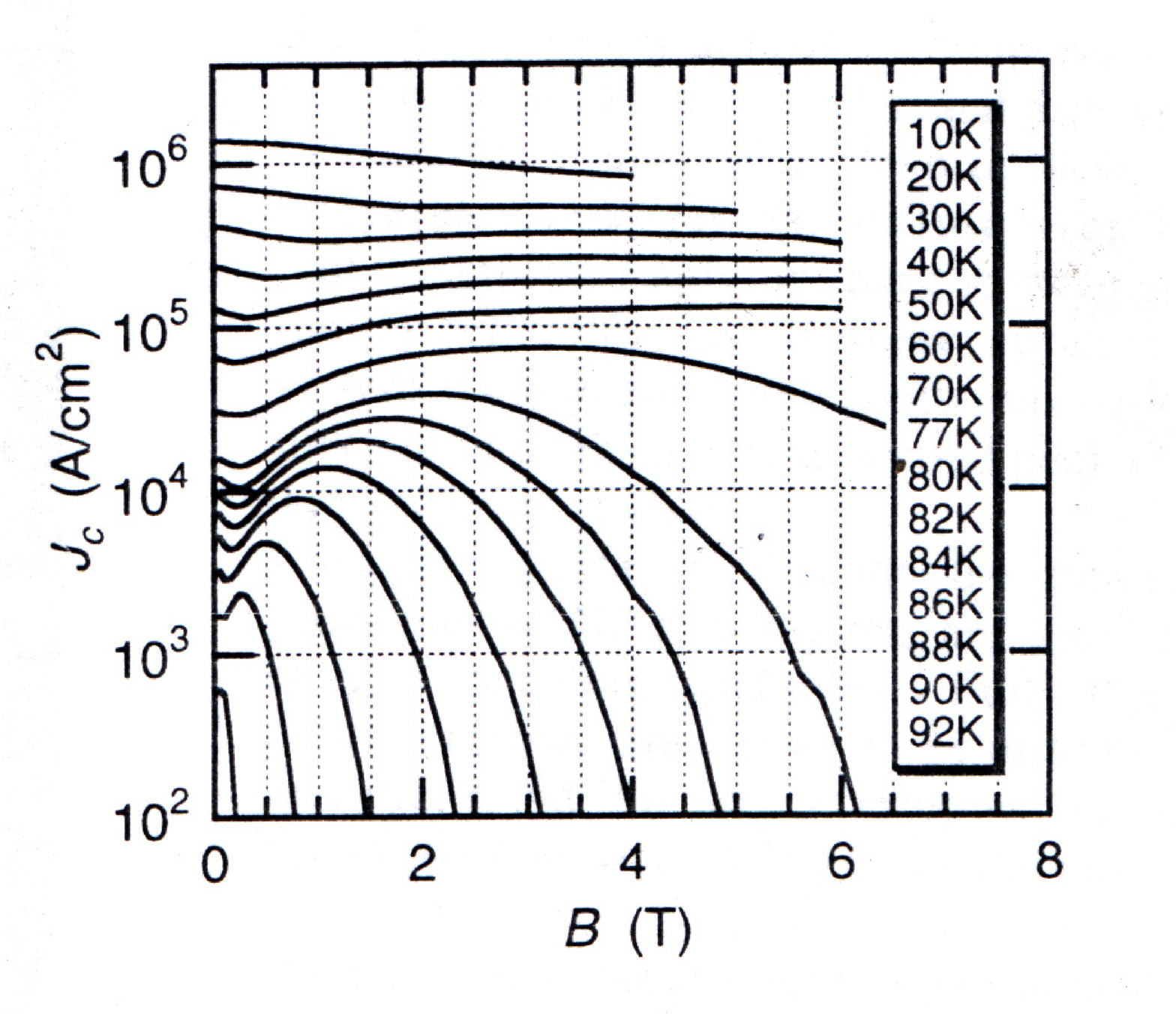}}
 \vspace{.1cm}
\caption{The experimental data of the $J_{c}-B$ curves which is in
Semi-logarithmic form \protect\cite{3}.}
\end{figure}

\begin{figure}[F2]
\epsfxsize= .83\hsize  \vskip 1.0\baselineskip \centerline{
\epsffile{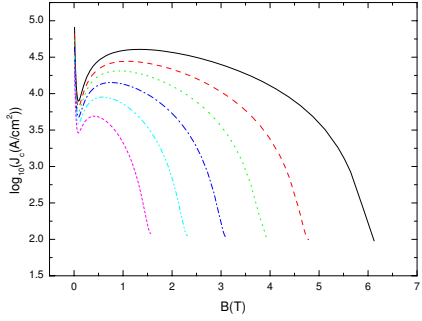}}
 \vspace{.1cm}
\caption{The numerical solutions of Eq. (12) with $3\nu =2$, $l=0.68$, $%
m=9.8$.}
\end{figure}

\begin{figure}[F3]
\epsfxsize= .83\hsize  \vskip 1.0\baselineskip \centerline{
\epsffile{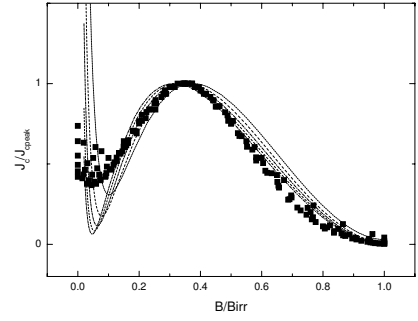}}
 \vspace{.1cm}
\caption{The scaling behavior of peak effect with $B/B_{irr}$ and $%
J_c/J_{cpeak}$ . The lines are data from Fig.1 and the open symbols are
datas from Fig.2}
\end{figure}

\end{multicols}

\end{document}